\documentclass[a4paper,11pt]{article}
\usepackage{pos}

\title{\texorpdfstring{$t\bar{t}b\bar{b}$}{ttbb} predictions at NLO in QCD and \texorpdfstring{$b$}{b}-jet modelling}

\author*[a]{Michele Lupattelli}

\affiliation[a]{Institute for Theoretical Particle Physics and Cosmology,\\
  RWTH Aachen University, D-52056 Aachen, Germany}

\emailAdd{lupattelli@physik.rwth-aachen.de}

\abstract{In this contribution we report on the latest theoretical predictions for $t\bar{t}b\bar{b}$. These results are obtained for the fully decayed final state $pp \rightarrow e^+ \nu_e \mu^- \nu_{\mu} b \bar{b} b \bar{b}$ at next-to-leading order accuracy in QCD. We present predictions with full off-shell effects and in the Narrow Width Approximation. Specifically, we report on the size of the NLO QCD corrections and the size of the full off-shell effects. Finally, we study the nature of the $b$-jets present in the final state.}

\FullConference{%
  41st International Conference on High Energy physics - ICHEP2022\\
  6-13 July, 2022\\
  Bologna, Italy
}

\begin{document}

\maketitle


\section{Introduction}
One of the greatest achievements of the LHC is the discovery of the Higgs boson in 2012. The properties of this newly discovered particle have been tested ever since. For instance, to better understand the fundamental interactions, it is crucial to investigate how the Higgs boson couples to the heaviest of the quarks, the top quark. Direct information on the top-Yukawa coupling can be extracted from the production process of a Higgs boson in association with a top-quark pair $t\bar{t}H$. This is a rare process, representing only 1\% of the total Higgs-boson production rate. However, it has finally been observed in 2018 by both the ATLAS \cite{ATLAS:2018mme} and CMS \cite{CMS:2018uxb} collaborations. We can look for this process in several decay channels of the Higgs boson. The $H\rightarrow b\bar{b}$ decay channel has the largest branching ratio ($\mathcal{BR}=58\%$) and is therefore suitable for the collection of a larger amount of data. On the other hand, this final state presents many jets (at least four $b$-jets) and thus suffers from a huge background. As a consequence, to investigate $t\bar{t}H(H\rightarrow b \bar{b})$ we first need to have a proper description of the background processes. One of many backgrounds is the direct production of $t\bar{t}b\bar{b}$, which has been measured by both the ATLAS \cite{ATLAS:2018fwl} and CMS \cite{CMS:2020grm} collaborations.

In this contribution I will report on the latest theoretical predictions on $t\bar{t}b\bar{b}$. They have been obtained at next-to-leading order (NLO) accuracy in QCD in the dileptonic decay channel of the top quarks using the \textsc{Helac-Nlo} Monte-Carlo framework \cite{Bevilacqua:2011xh}. In Ref. \cite{Bevilacqua:2021cit} the calculation is performed including full off-sheel effects, i.e. the unstable particles are described by Breit-Wigner propagators and all double-, single- and non-resonant top-quark contributions as well as all interference effects are consistently incorporated at the matrix element level. There we also studied the size of the NLO QCD corrections, which are reported in Sec. \ref{sizeNLO}. In Ref. \cite{Bevilacqua:2022twl} we performed the calculation for the same process in the Narrow Width Approximation (NWA). The calculation is
considerably simplified compared to the full off-shell one since the heavy resonances are now described by Dirac deltas that force the on-shell production of unstable particles. This also yields a factorization of the cross section into production times decays. Moreover, only double-resonant top-quark diagrams are considered. We then compared the NWA calculation to the full off-shell one to investigate the size of the full off-shell effects, as we report in Sec.\ref{sizeOffSh}. Finally, we exploited the properties of the NWA to study the nature of the $b$-jets present in the final state, as we outline in Sec. \ref{bjet}.

\section{Size of NLO QCD corrections}
\label{sizeNLO}

In Ref. \cite{Bevilacqua:2021cit} we computed the NLO QCD corrections to the fully decayed final state $pp \rightarrow e^+ \nu_e \mu^- \nu_{\mu} b \bar{b} b \bar{b}$ considering full off-shell effects. We performed the calculation using rather inclusive cuts
\begin{equation}
    p_T(\ell) > 20~\text{GeV}, \qquad p_T(b) > 25~\text{GeV}, \qquad
    |\eta(\ell)|<2.5, \qquad |\eta(b)| < 2.5,
\end{equation}
where $\ell=e^+,\mu^-$. The leading order (LO) and NLO predictions for the integrated fiducial cross section are
\begin{equation}
    \label{fidXS}
    \sigma^{LO} = 6.813^{+4.338(64\%)}_{-2.481(36\%)}~\text{fb}, \qquad
    \sigma^{NLO} = 13.22^{+2.65(20\%)}_{-2.96(22\%)}~\text{fb}.
\end{equation}
These results were obtained using the \texttt{NNPDF3.1} PDF set and the dynamical scale choice for the renormalization and factorization scales $\mu_R = \mu_F = \mu_0 = H_T/3$, where
\begin{equation}
    H_T= p_T(b_1) + p_T(b_2) + p_T(b_3) + p_T(b_4) + p_T(e^+) + p_T(\mu^-) + p_T^{\text{miss}}.
\end{equation}
We estimated the uncertainty on the predictions using the standard $7$-point scale variation, meaning that we varied separately the renormalization and factorization scales according to
\begin{equation}
    \biggl ( \frac{\mu_R}{\mu_0},\frac{\mu_F}{\mu_0} \biggr ) = \bigl \{ (0.5,0.5), (1,0.5), (0.5,1), (1,1), (1,2), (2,1), (2,2) \bigr \}.
\end{equation}
The PDF uncertainty has proved to be negligible compared to the scale uncertainty ($\sim 1\%$ at NLO). From the results of Eq. \eqref{fidXS} we can see that the NLO QCD corrections are very important for this process, about $94\%$. Moreover, the theoretical uncertainty is dramatically reduced when we include those corrections, going from $64\%$ to $22\%$.

\begin{figure}[t]
  \begin{center}
  \includegraphics[width=0.5\textwidth]{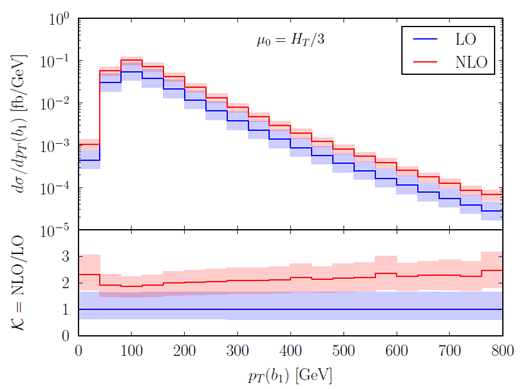}
  \includegraphics[width=0.4\textwidth]{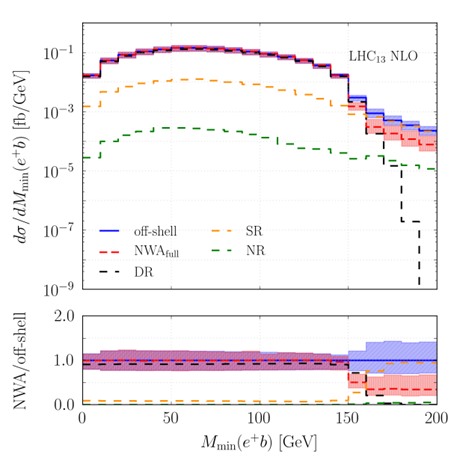}
\end{center}
\caption{\label{fig:NLOeff_offsheff} 
\it Differential cross-section distributions as a function of
$p_T(b_1)$ (left) and as a function of $M_{\text{min}}(e^+ b)$ (right). On the left plot the LO prediction (blue) is compared to the NLO one (red). On the right plot the NWA prediction (red) is compared to the full off-shell prediction (blue). Moreover, the double- (black), single- (yellow) and non-resonant (green) top-quark contributions to the full off-shell prediction are reported. Figures taken from Ref. \cite{Bevilacqua:2021cit} (left) and Ref. \cite{Bevilacqua:2022twl} (right). }
 \end{figure}

We also investigated several differential distributions. For instance, in Fig. \ref{fig:NLOeff_offsheff} (left) we present the differential cross-section as a function of the hardest $b$-jet $p_T(b_1)$. From this observable we can see that also at the differential level the NLO QCD corrections are very important and they introduce a shape distortion, with the $K$-factor ranging from $90\%$ to $135\%$.

\section{Size of full off-shell effects}
\label{sizeOffSh}

In Ref. \cite{Bevilacqua:2022twl} we investigated further aspects of the $t\bar{t}b\bar{b}$ process. To this end, we repeated the calculation outlined in the previous section, this time using the NWA. Comparing these new results to the full off-shell ones reported in Ref. \cite{Bevilacqua:2021cit} we were able to extract the size of the full off-shell effects. One should expect these effects to be relevant only in specific phase-space regions, such as kinematic edges and high energy tales of differential distributions. Indeed, the NWA is very good in describing the integrated fiducial cross section
\begin{equation}
    \sigma^{NLO}_{\text{Off-shell}} = 13.22^{+2.65(20\%)}_{-2.96(22\%)}~\text{fb}, \qquad
    \sigma^{NLO}_{\text{NWA}} = 13.16^{+2.61(20\%)}_{-2.93(22\%)}~\text{fb}.
\end{equation}
We can indeed see that the two predictions are in excellent agreement, with the off-shell effects of the order of $\sim 0.5\%$.

We also investigated several differential distributions. For most of the observables, the off-shell effects turned out to be negligible in the energy ranges we considered. On the other hand, the off-shell effects become really important for observables with kinematic edges. Here we report on the $M_{\text{min}}(e^+ b)$ observable. This observable has a kinematic edge at the value $\sqrt{m_t^2-m_W^2}\approx 153$ GeV. This kinematic edge is sharp for the LO NWA calculation, while it is smeared out by the extra radiation at NLO. Moreover, the full off-shell calculation presents also single- and non- resonant top-quark contributions, which further contribute to this smearing. We can see from Fig. \ref{fig:NLOeff_offsheff} (right) that the NWA prediction does not describe properly this observable beyond the kinematic edge, as we would expect. Indeed, there the single-resonant top-quark contribution to the full off-shell prediction (yellow line), which is completely absent in the NWA case, becomes dominant. The differences between the full off-shell and the NWA are large and in the range we investigated they can reach $66\%$.

\section{\texorpdfstring{$b$}{b}-jet labelling}
\label{bjet}

\begin{figure}[t]
  \begin{center}
  \includegraphics[width=0.35\textwidth]{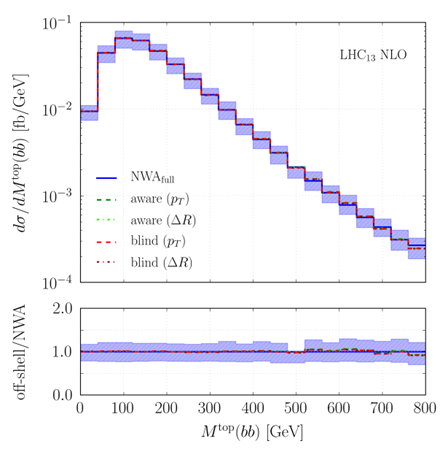}
  \includegraphics[width=0.35\textwidth]{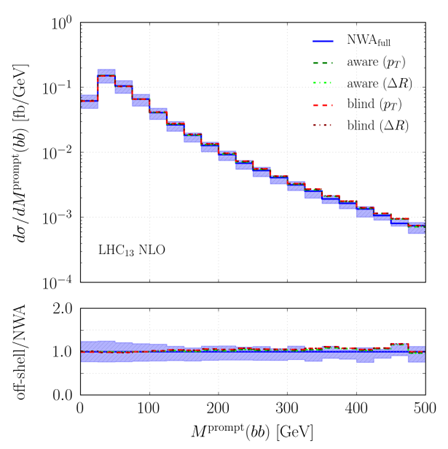}
\end{center}
\caption{\label{fig:blab} 
\it Differential cross-section distributions as a function of
$M^{\text{top}}(bb)$ (left) and $M^{\text{prompt}}(bb)$ (right). Several versions of our labelling prescription are compared to the reference NWA in blue. Figures taken from Ref. \cite{Bevilacqua:2021cit} }
 \end{figure}

We define two categories of $b$-jets in $t\bar{t}b\bar{b}$:
\begin{itemize}
    \item \textit{$b$-jets from top-quark decays}: the decay products of the top quark and antiquark;
    \item \textit{prompt $b$-jets}: extra $b$-jet pair, mainly produced in gluon splitting.
\end{itemize}
In Ref. \cite{Bevilacqua:2022twl} we exploited the factorization property of the NWA cross section into production times decay to investigate the nature of the $b$-jets present in the $t\bar{t}b\bar{b}$ process. Indeed, this property allows us to label the $b$-jets prior the event generation. This is of course unrealistic, since in data analysis we cannot access such information. However, this can be really useful for comparison purposes. Thus, we proposed a prescription to distinguish the prompt $b$-jets from the $b$-jets from top-quark decays and we used it to provide predictions in the full off-shell case, where we cannot access the information on the origin of the $b$-jets present in the final state. To this end, we reconstruct all the possible resonant histories that can give rise to the final state and we choose the one that minimizes the function
\begin{equation}
    \label{Qfunc}
    Q=|M(t)-m_t|\times|M(\bar{t})-m_t|\times M(bb).
\end{equation}
With this function we are constraining the mass of the reconstructed top quarks to be as close as possible to its nominal mass, and the invariant mass of the prompt $b$-jets to zero, since they are mainly produced in gluon splittings. In Fig. \ref{fig:blab} we report the invariant mass distributions of the $b$-jet pair from top-quark decays $M^\text{top}(bb)$ and of the prompt $b$-jet pair $M^\text{prompt}(bb)$. We compared several versions of our prescription to the reference NWA predictions. In detail, we either considered (aware) or not (blind) the charge of the $b$-jets in the resonant history reconstruction. Moreover, the gluon splitting can occur in the top-quark decay, i.e. we can have the decay process $t\rightarrow W^+ b b \bar{b}$: in this case, we either choose the $b$-jet from top-quark decay to be the one with the highest transverse momentum $p_T$, or the one that does not minimize $\Delta R(b\bar{b})$. From Fig. \ref{fig:blab} we can see that the difference between the various versions is negligible, and that the full off-shell prediction obtained using Eq. \eqref{Qfunc} are in excellent agreement with the reference NWA.

\begin{figure}[t]
  \begin{center}
  \includegraphics[width=0.3\textwidth]{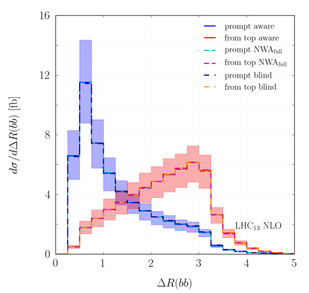}
  \includegraphics[width=0.3\textwidth]{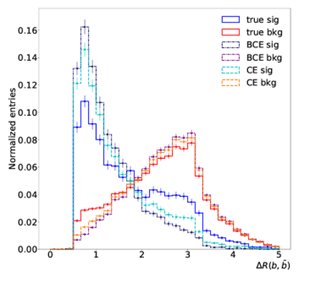}
  \includegraphics[width=0.3\textwidth]{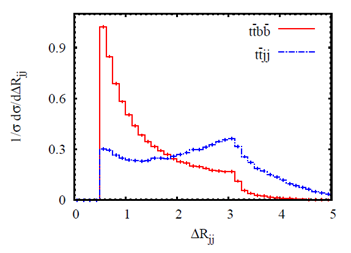}
\end{center}
\caption{\label{fig:comparison} 
\it Differential cross-section distributions as a function of
$\Delta R^{\text{top}}(bb)$ and $\Delta R^{\text{prompt}}(bb)$. The predictions on the left are obtained using our labelling prescription, the central ones using machine learning techniques and, finally, the right ones are predictions for the undecayed final state $pp\rightarrow t\bar{t}b\bar{b}$, meaning that the top quarks are treated as stable particles. For the plot on the right, only the red curve refers to the $t\bar{t}b\bar{b}$ process. Specifically, it is the $\Delta R^{\text{prompt}}(bb)$ distribution. Figures taken from Ref. \cite{Bevilacqua:2022twl, Jang:2021eph, Bevilacqua:2014qfa}. }
 \end{figure}

To support our results, we compared the shape of our predictions for invariant mass distributions $M(bb)$ and angular separations $\Delta R(bb)$ to distributions for the same observables obtained with machine learning techniques \cite{Jang:2021eph} and with stable top quarks \cite{Bevilacqua:2014qfa}. In the latter case we can compared only the prompt $b$-jet pair distributions, since the top quarks do not decay. As we can see from Fig. \ref{fig:comparison}, very good agreement is found in the shape of the distributions as well as in the location of the peaks. This confirms even further that our simple $b$-jet labelling prescription works very well.

\section{Summary}

In this contribution we presented results for $pp \rightarrow e^+ \nu_e \mu^- \nu_{\mu} b \bar{b} b \bar{b}$ both with full off-shell effects and in the NWA. The NLO QCD corrections to this process are very important, around $94\%$ at the integrated fiducial level, and can be even larger at the differential level. Moreover, the theoretical uncertainty is sizeable, around $20\%$ at NLO. This suggests that NNLO corrections are needed for this process. We then compared the NWA predictions to the full off-shell ones and found that the full off-shell effects become important in the tails of distributions with kinematic edges. Finally, we proposed a prescription to label the $b$-jets in the final state.

\section{Acknowledgments}

The research of M. Lupattelli was supported by the DFG under grant 400140256 - GRK 2497: \textit{The physics of the heaviest particles at the Large Hardon Collider}.

\bibliographystyle{JHEP}
\bibliography{references} 

\providecommand{\href}[2]{#2}\begingroup\raggedright\begin{thebibliography}{1}

\bibitem{ATLAS:2018mme}
{\bf ATLAS} Collaboration, M.~Aaboud et~al., {\it {Observation of Higgs boson
  production in association with a top quark pair at the LHC with the ATLAS
  detector}},  {\em Phys. Lett. B} {\bf 784} (2018) 173--191,
  [\href{http://arxiv.org/abs/1806.00425}{{\tt arXiv:1806.00425}}].

\bibitem{CMS:2018uxb}
{\bf CMS} Collaboration, A.~M. Sirunyan et~al., {\it {Observation of
  $\mathrm{t\overline{t}}$H production}},  {\em Phys. Rev. Lett.} {\bf 120}
  (2018), no.~23 231801, [\href{http://arxiv.org/abs/1804.02610}{{\tt
  arXiv:1804.02610}}].

\bibitem{ATLAS:2018fwl}
{\bf ATLAS} Collaboration, M.~Aaboud et~al., {\it {Measurements of inclusive
  and differential fiducial cross-sections of $ t\overline{t} $ production with
  additional heavy-flavour jets in proton-proton collisions at $ \sqrt{s} $ =
  13 TeV with the ATLAS detector}},  {\em JHEP} {\bf 04} (2019) 046,
  [\href{http://arxiv.org/abs/1811.12113}{{\tt arXiv:1811.12113}}].

\bibitem{CMS:2020grm}
{\bf CMS} Collaboration, A.~M. Sirunyan et~al., {\it {Measurement of the cross
  section for $\text{t}\bar{\text{t}}$ production with additional jets and b
  jets in pp collisions at $\sqrt{s}=$ 13 TeV}},  {\em JHEP} {\bf 07} (2020)
  125, [\href{http://arxiv.org/abs/2003.06467}{{\tt arXiv:2003.06467}}].

\bibitem{Bevilacqua:2011xh}
G.~Bevilacqua, M.~Czakon, M.~V. Garzelli, A.~van Hameren, A.~Kardos, C.~G.
  Papadopoulos, R.~Pittau, and M.~Worek, {\it {HELAC-NLO}},  {\em Comput. Phys.
  Commun.} {\bf 184} (2013) 986--997,
  [\href{http://arxiv.org/abs/1110.1499}{{\tt arXiv:1110.1499}}].

\bibitem{Bevilacqua:2021cit}
G.~Bevilacqua, H.-Y. Bi, H.~B. Hartanto, M.~Kraus, M.~Lupattelli, and M.~Worek,
  {\it {$ t\overline{t}b\overline{b} $ at the LHC: on the size of corrections
  and b-jet definitions}},  {\em JHEP} {\bf 08} (2021) 008,
  [\href{http://arxiv.org/abs/2105.08404}{{\tt arXiv:2105.08404}}].

\bibitem{Bevilacqua:2022twl}
G.~Bevilacqua, H.-Y. Bi, H.~B. Hartanto, M.~Kraus, M.~Lupattelli, and M.~Worek,
  {\it {$t\bar{t}b\bar{b}$ at the LHC: On the size of off-shell effects and
  prompt $b$-jet identification}},  [\href{http://arxiv.org/abs/2202.11186}{{\tt
  arXiv:2202.11186}}].

\bibitem{Jang:2021eph}
C.~Jang, S.-K. Ko, J.~Choi, J.~Lim, Y.-K. Noh, and T.~J. Kim, {\it {Learning to
  increase matching efficiency in identifying additional b-jets in the $\text
  {t}\bar{\text {t}}\text {b}\bar{\text {b}}$ process}},  {\em Eur. Phys. J.
  Plus} {\bf 137} (2022), no.~7 870,
  [\href{http://arxiv.org/abs/2103.09129}{{\tt arXiv:2103.09129}}].

\bibitem{Bevilacqua:2014qfa}
G.~Bevilacqua and M.~Worek, {\it {On the ratio of $ t\overline{t} b\overline{b}
  $ and $ t\overline{t} jj $ cross sections at the CERN Large Hadron
  Collider}},  {\em JHEP} {\bf 07} (2014) 135,
  [\href{http://arxiv.org/abs/1403.2046}{{\tt arXiv:1403.2046}}].

\end{thebibliography}\endgroup


\end{document}